\documentclass[11pt]{article}
\usepackage[margin=1in]{geometry}
\usepackage{amssymb,amsfonts,amsmath,amsthm,amscd,dsfont,mathrsfs}
\usepackage{graphicx,float,psfrag,epsfig,color}

\newcommand{\iid}{\stackrel{\text{iid}}{\sim}}
\newcommand{\mR}{\mathbb{R}}
\newcommand{\mN}{\mathbb{N}}
\newcommand{\mZ}{\mathbb{Z}}
\newcommand{\mC}{\mathbb{C}}

\newcommand{\tr}{\mathrm{tr}}

\newcommand{\pp}{\mathbb{P}}

\DeclareMathOperator{\diag}{diag}
\newcommand{\x}{x^*}
\newcommand{\p}{p^*}

\theoremstyle{definition}
\theoremstyle{plain}
\theoremstyle{plain}
\theoremstyle{plain}
\newtheorem{lemma}{Lemma}%[section]
\theoremstyle{plain}
\theoremstyle{plain}
\theoremstyle{plain}
\newtheorem{conjecture}{Conjecture}%[section]
\theoremstyle{remark}
\begin{document}

\title{Proof of the outage probability conjecture for MISO channels}
\author{Emmanuel Abbe, Shao-Lun Huang, Emre Telatar}
\date{}
\maketitle

\begin{abstract}
In Telatar 1999 (\cite{telatar}), it is conjectured that the covariance matrices minimizing the outage probability for MIMO channels with Gaussian fading are diagonal with either zeros or constant values on the diagonal. In the MISO setting, this is equivalent to conjecture that the Gaussian quadratic forms having largest tale probability correspond to such diagonal matrices.  
%The conjecture can equivalently be stated in terms of finding optimal weights for a weighted sum of exponential independent random variables.
We prove here the conjecture in the MISO setting. 
%Some extensions and applications of the conjecture are also discussed.  
\end{abstract}

\section{The conjecture}

\begin{figure}
\begin{center}
\includegraphics[scale=.72]{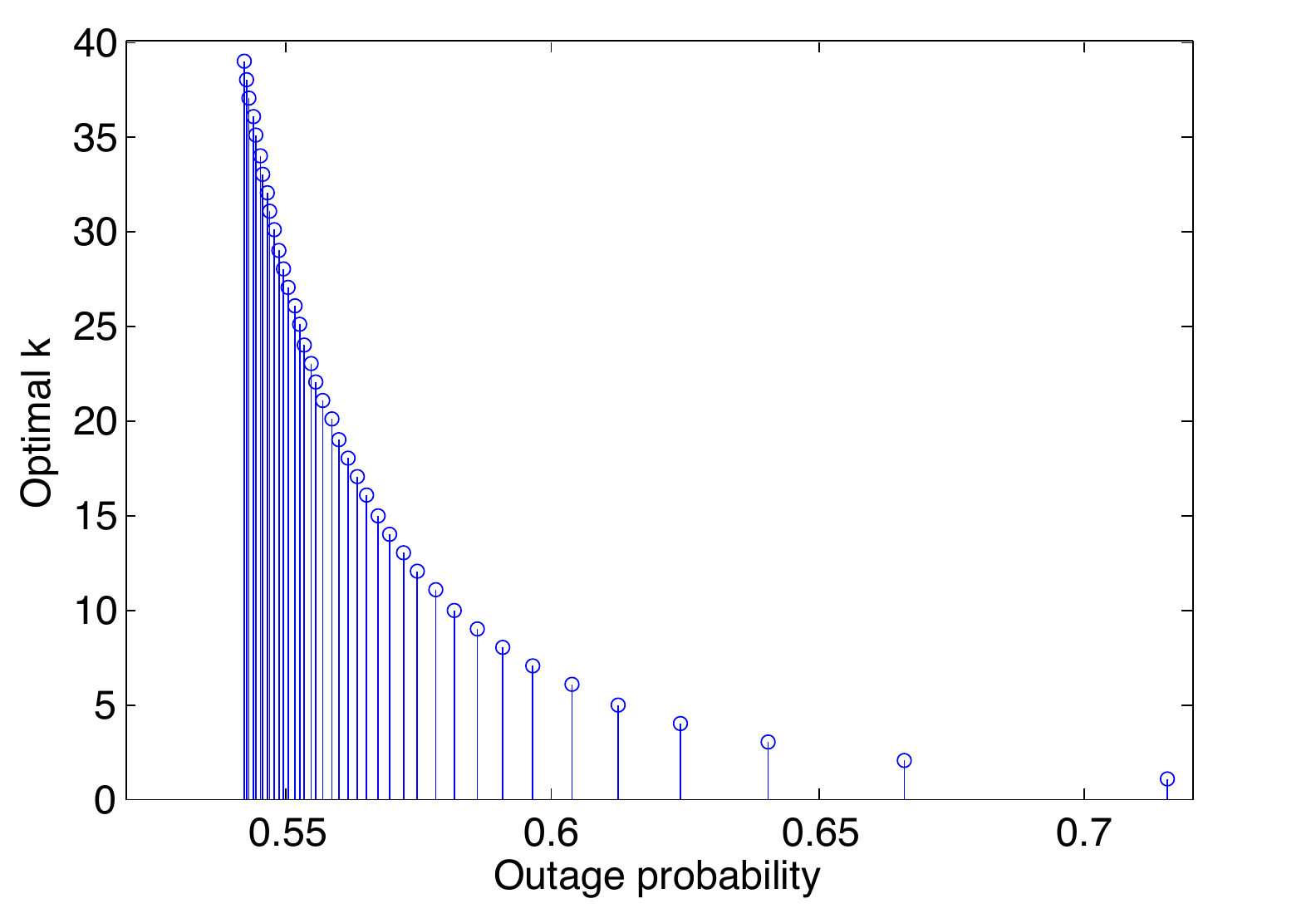}
\caption{Plots of the optimal value of $k$ (the minimizer of \eqref{obj}) for a given value of $\min_{Q \in \xi(t)} \pp \{H Q H^* \leq x\}$ (outage probability) when $t=40$. A tab indicates the first location at which the optimal $k$ decreases from one, in particular, before the first tab, the optimal $k$ is equal to $t=40$. Note that the locations of the tabs do not depend on $t$; however, for $t$ larger than 40, more tabs will appear before the first tab plotted here.}
\label{t16}
\end{center}
\end{figure}

\begin{conjecture}\label{angle}
Let  $\xi(t):= \{  Q \in \mC^{t\times t} \mid Q \geq 0, \, \tr Q \leq 1\}$ and ${(H_i)}_{1\leq i \leq t} \iid \mathcal{N}_{\mC}(1)$.
%and let $||H||_Q^2=  \langle H,H\rangle_Q=HQH^*$.
For all $x \in \mR$, there exists $k \in \{1,\ldots,t\}$ such that 
\begin{align}
 \arg\min_{Q \in \xi(t)} \pp \{H Q H^* \leq x\} = \{ U \diag (\underbrace{\frac{1}{k},\ldots,\frac{1}{k}}_k,0,\ldots,0) U^* : U \in U(t) \}, \label{obj}
 \end{align}
where $U(t)$ denotes the group of $t \times t$ unitary matrices.
\end{conjecture}
Note that the relation between $k$ and $x$ relies on properties of Gamma distributions, and one can easily compute for a given $x$ the optimal value of $k$. Figure \ref{t16} shows the value of the optimal $k$ for a given outage probability (i.e., for a given $\min_{Q \in \xi(t)} \pp \{H Q H^* \leq x\}$) when $t=40$. The lowest value of the outage probability for which the number of active antennas is strictly less than $t$ (i.e., when the optimal $Q$ has a zero entry) is slightly above half\footnote{Although this is a fairly high outage probability, there may be schemes (such ARQ schemes) that may operate at such high outage probability regimes at the beginning of a communication.}.  

This conjecture can be interpreted in different ways. It characterizes the optimal power allocation over the antennas of a non-ergodic MISO channel with Gaussian fading, in order to minimize the outage probability. Geometrically, it characterizes the best choice of a norm induced by a positive definite matrix, to minimize the probability of observing a short random vector which is Gaussian distributed. 

Observe that one can w.l.o.g.\ consider diagonal matrices $Q$ (since $H$ is unitary invariant), and the conjecture can also be expressed as follows.

%\begin{conj}
%Let  $n \in \mN^*$, $Q \in \xi(n):= \{M \in \mC^{n \times n}| M \text{ p.s.d., }\tr(M)\leq 1 \}$, $H=(H_1,\ldots,H_n)$ with ${(H_i)}_{1\leq i \leq n} \stackrel{iid}{\sim} \mathcal{N}_{\mC}(1)$ and $\langle H,H\rangle_Q:=HQH^*$. For all $x \in \mR$, $\exists k \in \{1,\ldots,n\}$ s.t.
%$$ \arg\min_{Q \in \xi(n)} \pp \{ \langle H,H\rangle_Q \leq x\} \ni \diag (\underbrace{\frac{1}{k},\ldots,\frac{1}{k}}_k,0,\ldots,0)  .$$
%\end{conj}
%\vspace{1.1cm}

%Note that the following is an equivalent formulation of this conjecture.
\begin{conjecture}\label{miso}
Let $\theta(t):=\{ x \in \mR_+^t | \sum_{i=1}^t x_i \leq1\}$, $X=(X_1,\ldots,X_t)$ with ${\{X_i\}}_{1\leq i \leq t} \iid \mathcal{E}(1)$ and $\langle q,X \rangle:= \sum_{i=1}^t q_i X_i$. For all $x \in \mR$, there exists $k \in \{1,\ldots,t\}$ s.t.
$$\arg\min_{q \in \theta(t)} \pp \{\langle q,X \rangle \leq x\} = \{  (\underbrace{\frac{1}{k},\ldots,\frac{1}{k}}_k,0,\ldots,0) \Pi : \Pi \in S(t) \},$$
where $S(t)$ denotes the group of $t \times t$ permutation matrices.
\end{conjecture}

This formulation also relates to portfolio optimization problems. In \cite{telatar}, the conjecture is stated in a slightly more general setting where $H$ is an $r \times t$ random matrix and where 
$$Q \mapsto \pp \{ \log \det(I_t + \mathrm{snr} H Q H^*) \leq R \}$$ is minimized.
%\begin{conjecture}\label{telatar}
%Let  $\xi(t):= \{  Q \in \mC^{t\times t} \mid Q \geq 0, \, \tr Q \leq 1\}$ and ${(H_{i,j})}_{1\leq i \leq r, 1\leq j \leq t} \iid \mathcal{N}_{\mC}(1)$.
%%and let $||H||_Q^2=  \langle H,H\rangle_Q=HQH^*$.
%For all $R \in \mR$, there exists $k \in \{1,\ldots,t\}$ s.t.
%$$ \arg\min_{Q \in \xi(t)} \pp \{ \log \det(I_t + H Q H^*) \leq R \} =\{ U \diag (\underbrace{\frac{1}{k},\ldots,\frac{1}{k}}_k,0,\ldots,0) U^* : U \in U(t)\}. $$
%\end{conjecture}
Clearly, if $r=1$, this is equivalent to the above conjecture, with $x=\frac{e^R-1}{\mathrm{snr}}$. Many works in the literature about MIMO channels that investigate properties of the outage probability assume that the conjecture holds or pick a uniform power allocation without discussing its optimality. In \cite{abbe_thesis}, Conjecture \ref{angle} is proved for low values of $t$. We complete here the proof for an arbitrary $t$.  
%We now present a proof of Conjecture \ref{miso}. 

\section{Proof}
Recall that $X=(X_1,\ldots,X_t)$ with ${\{X_i\}}_{1\leq i \leq t} \iid \mathcal{E}(1)$. We denote by $f_Y$ the density of a random variable $Y$.

\begin{lemma}\label{lemma_trick}
Let $\widetilde{X}$ be such that $\widetilde{X} \stackrel{(d)}{\sim} X_1$ and $\widetilde{X},X$ are mutually independent. Then $\forall x \in \mR$, $\forall q \in (\mR_+^*)^t$ and $\forall k \in \{1,\ldots,t\}$,
\begin{eqnarray}
\frac{\partial \pp \{ \langle q,X \rangle \leq x\}}{\partial q_k} &=& -f_{\langle q,X \rangle+ q_k \widetilde{X}} (x) .\label{derivative}
\end{eqnarray}
\end{lemma}
\begin{proof}
As $f_{\langle q,X \rangle} \in \mathrm{L}^1(\mR) \cap \mathcal{C}^0(\mR_+^*)$ (only when $t=1$ there is a discontinuity at $x=0$), we can use the Fourier transform to write:
\begin{eqnarray*}
 f_{\langle q,X \rangle} (x)&=& \frac{1}{2\pi} \int_{\mR}  \prod_{j=1}^t {(1+ \omega i q_j)}^{-1}  e^{\omega i x} d\omega, \,\,\, \forall x \in \mR_+^*,
\end{eqnarray*}
therefore
\begin{eqnarray}
 \pp \{ \langle q,X \rangle \leq x\}
&=&  \frac{1}{2\pi} \int_{\mR}\prod_{j=1}^t {(1+\omega i q_j)}^{-1} \frac{1}{\omega i} (e^{\omega ix}-1)  d\omega, \,\,\, \forall x \in \mR_+^* \label{Pout}
\end{eqnarray}
and is zero for negative values of $x$. Thus we get
\begin{eqnarray*}
\frac{\partial   \pp \{ \langle q,X \rangle \leq x\}}{\partial q_k} &=& - \frac{1}{2\pi} \int_{\mR} \prod_{j=1}^t {(1+ \omega i q_j)}^{-1} {(1+ \omega i q_k)}^{-1}  e^{\omega i x} d\omega
 \\
&=&  -f_{\langle q,X \rangle+ q_k \widetilde{X} } (x).
\end{eqnarray*}
%where
%$\widetilde{X}$ is such that $\widetilde{X}$ and $X$ are mutually independent and $\widetilde{X} \stackrel{(d)}{\sim} X_1$.
\end{proof}
\begin{lemma}\label{lemma_trick2}
Let $Y$ be a random variable independent of $X_1,X_2 \stackrel{iid}{\sim} \mathcal{E}(1)$,
and let $x,q_1,q_2\in \mR$. We then have
\begin{eqnarray}
f_{Y+ q_1 X_1} (x) -f_{Y+ q_2 X_2} (x) &=& (q_2 -q_1)f^{'}_{Y+ q_1 X_1+q_2 X_2}(x)   \label{trick2}
\end{eqnarray}
\end{lemma}
This is easily verified by using the Fourier transform.
\begin{lemma}\label{unimodal}
For all $t \geq 2$ and $q \in (\mR_+^*)^t$, we have $f_{\langle q,X \rangle} \in \mathcal{C}^{\infty}(\mR^*) \cap \mathcal{C}^{t-2}(\mR)$ and there exists a unique $a \in \mR^*_+$ such that $f_{\langle q,X \rangle}^{'} (x) > 0$, $\forall 0<x < a$, $f_{\langle q,X \rangle}^{'} (a) = 0$ and $f_{\langle q,X \rangle}^{'} (x) < 0$, $ \forall x > a$.
%Moreover, if
% $f_{\langle q,X \rangle}^{'} (x)=0$ for $x \in \mR_+^*$, then $f_{\langle q,X \rangle}^{''} (x) \neq 0$.
\end{lemma}
\begin{proof}
The fact that $f_{\langle q,X \rangle} \in \mathcal{C}^{\infty}(\mR^*) \cap \mathcal{C}^{t-2}(\mR)$ can be verified by induction, knowing that the exponential density is in $\mathcal{C}^{\infty}(\mR^*)$ and using properties of convolution and differentiation.

\noindent
Note that $\forall t \geq 2$ and for $q \in \theta(t)$ s.t. all $q_i$'s are different, which can be written without loss of generality as $0<q_1 < q_2< \dots<q_t$, we have
\begin{eqnarray}
f_{\langle q,X \rangle}(x) = \sum_{i=1}^t \prod_{j\in \{1,\dots,t\} \atop \text{s.t.} i \neq j} \frac{1}{q_i -q_j} {q_i}^{t-1}  f_{q_i X_i} (x), \quad \forall x \in \mR. \label{convol}
\end{eqnarray}
This can also be verified by induction. Moreover, we have that $\forall x \in \mR, \forall t\geq 2$ the function $$q  \mapsto \sum_{i=1}^t \prod_{j\in \{1,\dots,t\} \atop \text{s.t.} i \neq j} \frac{1}{q_i -q_j} {q_i}^{t-1}  f_{q_i X_i} (x) \in \mR_+$$ is continuous and \eqref{convol} converges when considering equal $q_i$'s. So we can restrict ourself to prove the lemma for $q$'s having all components different (and we will consider such $q$'s in what follows).

\noindent
For $t\geq 2$, we have $f_{\langle q,X \rangle}^{(k)} (0) = 0$, $\forall k=0,\ldots,t-2$ (this is a consequence of the first statement in the lemma). Let us suppose that there exist $a,b>0$ such that $a \neq b$ and
\begin{eqnarray}
f_{\langle q,X \rangle}^{'} (a) =f_{\langle q,X \rangle}^{'} (b) = 0. \label{hyp}
\end{eqnarray}
From \eqref{convol}, the assumption \eqref{hyp}, in addition with  $f_{\langle q,X \rangle}^{(k)} (0) = 0$ for $k=1,\ldots,t-2$, implies that there exist $\alpha_1,\ldots,\alpha_t \in \mR$ and $\beta_1,\ldots,\beta_t \in \mR$, all different and non-zero, such that
$$\begin{pmatrix}
     1 &\cdots&1    \\
      \beta_1& \cdots&\beta_t\\
      \vdots &&\vdots\\
      \beta_1^{t-3}& \cdots&\beta_t^{t-3}\\
      e^{a \beta_1}& \cdots&e^{a \beta_t}\\
      e^{b \beta_1}& \cdots&e^{b \beta_t}
\end{pmatrix}
\begin{pmatrix}
      \alpha_1  \\ \vdots \\ \alpha_t
\end{pmatrix}
=0.$$
%without using it now, but keeping in mind that we also could have add
%$$\begin{pmatrix}
 %     \beta_1^{n-2}& \cdots&\beta_n^{n-2}
%\end{pmatrix}
%\begin{pmatrix}
 %      \alpha_1  \\ \vdots \\ \alpha_n
%\end{pmatrix} .$$
%But if
%$$\det \begin{pmatrix}
 %    1 &\cdots&1    \\
 %     \beta_1& \cdots&\beta_n\\
 %     \vdots &&\vdots\\
 %     \beta_1^{n-3}& \cdots&\beta_n^{n-3}\\
 %     e^{-a \beta_1}& \cdots&e^{-a \beta_n}\\
%      e^{-b \beta_1}& \cdots&e^{-b \beta_n}
%\end{pmatrix}
%=0,
%$$
But this is to say that there exists $a_0,\ldots,a_{t-1} \in \mR$, non all equal to zero, and $c \in \mR_+^*$ s.t.
\begin{eqnarray}
 \sum_{i=0}^{t-3} a_i x^i +a_{t-2}e^x+a_{t-1}e^{cx}=0, \,\,\, \forall x \in \{a \beta_1,\ldots,a \beta_t\}. \label{poly}
 \end{eqnarray}
Now, if $a_{t-1}=0$ and $a_{t-2}=0$, we clearly need  $a_0,\ldots,a_{t-3} =0$ to ensure $t$ solutions in \eqref{poly},
which leads to a contradiction.
 If $a_{t-1}=0$ or $a_{t-2}=0$, \eqref{poly} is equivalent to
\begin{eqnarray}
e^x = p(x), \label{exp}
\end{eqnarray}
where $p$ is a real polynomial of degree $t-3$.
Clearly, \eqref{exp} can have at most $t-2$ different solutions. Hence we have a contradiction with \eqref{poly}.
Otherwise, we have $a_{t-2}, a_{t-1} \neq 0$ and \eqref{poly} is equivalent to
\begin{eqnarray}
e^x + de^{cx} = \widetilde{p}(x), \label{exp2}
\end{eqnarray}
where $d \in \mR^*$ and $\widetilde{p}$ is a real polynomial of degree $t-3$.
Hence, \eqref{exp2} has at most $t-1$ different solutions, and we also have a contradiction. Thus, $a\neq b$.
The existence of $a$, as well as the sign of the derivatives around $a$ are easily verified. %This concludes the proof of the lemma.
%
%\noindent
%Let $x \in \mR_+^*$ such that
%$$f_{\langle q,X \rangle}^{'} (x)=0.$$
%%We then get from \eqref{trick2}
%%$$f_{\langle q,X \rangle}^{'} (x)=-1/q_1 f_{\langle q,X \rangle}(x)+ 1/q_1 f_{\langle q_{-1},X \rangle} (x)=0, $$
%%Therefore, $$f_{\langle q,X \rangle} (x)= f_{\langle q_{-1},X \rangle} (x).$$
%Let us assume (ab absurdo) that $$f_{\langle q,X \rangle}^{''} (x)=0.$$
%We then get from \eqref{trick2}
%$$f_{\langle q,X \rangle}^{''} (x)=-1/q_1 f_{\langle q,X \rangle}(x)^{'}+ 1/q_1 f_{\langle q_{-1},X_{-1} \rangle}^{'} (x)=0, $$
%where $q_{-1}= (q_2,\ldots,q_n) \in \theta(n-1)$ and $X_{-1}=(X_2,\ldots,X_n)$.
%Thus we get
%$$ f_{\langle q,X \rangle}^{'} (x)= f_{\langle q_{-1},X_{-1} \rangle}^{'} (x)=0. $$
%Last equality implies that $x$ is the unique point where both $f_{\langle q,X \rangle}$ and $f_{\langle q_{-1},X_{-1} \rangle}$ reach their maximum. But
%$$f_{\langle q,X \rangle}=f_{\langle q_{-1},X_{-1} \rangle} \star f_{q_1 X_1} $$
%and
%$$f_{\langle q,X \rangle}^{'}=f_{\langle q_{-1},X_{-1} \rangle}^{'} \star f_{q_1 X_1}, $$
%in particular
%$$ f_{\langle q,X \rangle}^{'} (x) = \int_0^x f_{\langle q_{-1},X_{-1} \rangle}^{'} (x-s) f_{q_1 X_1} (s) ds.$$
%By the first part of the lemma, we conclude that the last expression is non zero and thus we get a contradiction.
\end{proof}

\begin{proof}[Proof of conjecture \ref{miso}.]
Let $x \in \mR$. From lemma \ref{lemma_trick}, for any $0 \leq k \leq t$, $$\frac{\partial \pp \{ \langle q,X \rangle \leq x\}}{\partial q_k} = -f_{\langle q,X \rangle+ q_k \widetilde{X}_1} (x) \leq 0,$$
with $\widetilde{X}_1 $ independent of $X$ and $\widetilde{X}_1  \sim X_1$. We thus conclude that we can replace $\theta(t)$ by $\Theta(t):=\{ q \in \mR_+^t | \sum_{i=1}^t q_i =1\}.$
\\

\noindent
Using the Kuhn-Tucker theorem, if $q^* \in \Theta(t)$ minimizes $\pp \{ \langle q,X \rangle \leq x\}$, then $\exists \lambda \in \mR$ s.t.
\begin{eqnarray}
   \frac{\partial \pp \{ \langle q^*,X \rangle \leq x\} }{\partial q_k}
   \begin{cases}
     =\lambda,& \forall k \text{ s.t. } q_k^* > 0, \label{KT1} \\
     \geq \lambda, & \text{ otherwise}. \label{KT2}
   \end{cases}
\end{eqnarray}
By lemma \ref{lemma_trick} and \ref{lemma_trick2},
$$ \frac{\partial \pp \{ \langle q,X \rangle \leq x\}}{\partial q_k} = \frac{\partial \pp \{ \langle q,X \rangle \leq x\}}{\partial q_l} $$
is equivalent to
$$ (q_k-q_l)  f^{'}_{\langle q,X \rangle+ q_k \widetilde{X}_1+q_l \widetilde{X}_2} (x)=0$$
with $\widetilde{X}_1,\widetilde{X}_2,X$ mutually independent and $\widetilde{X}_2  \sim X_1$.
%This means that we cannot exclude the cases in which $q^*_k\neq q^*_l$ and still we do satisfy \eqref{KT1}.
Now, let us assume that $0 < q^*_1 < q^*_2 $ (this represent w.l.o.g. that at least two different non-zero values are in $q^*$). Then
\begin{eqnarray}
f^{'}_{\langle q^*,X \rangle+ q_1^* \widetilde{X}_1+q_2^* \widetilde{X}_2} (x)=0 \label{mod}
\end{eqnarray}
%and $$ f^{'}_{\langle q^*,X \rangle+ q_1^* \widetilde{X}_1+q_3^* \widetilde{X}_3} (x)=0, $$
%with $ \widetilde{X}_3$ such that $\widetilde{X}_1,\widetilde{X}_2,\widetilde{X}_3, X$ are mutually independent and $\widetilde{X}_3 \stackrel{(d)}{\sim} X_1$.
%We now will use \eqref{trick} from lemma \ref{lemma_trick} in several senses.\\
and using \eqref{trick2}, we get
\begin{eqnarray}
 f_{\langle q^*,X \rangle+ q_1^* \widetilde{X}_1+q_2^* \widetilde{X}_2} (x)= f_{\langle q^*,X \rangle+ q_1^* \widetilde{X}_1} (x) . \label{equ}
 \end{eqnarray}
We now assume that $q_3^*$, the third component of $q^*$, is non-zero. By successive use of \eqref{trick2} and by \eqref{equ}, we have
\begin{eqnarray}
&& f^{'}_{\langle q^*,X \rangle+ q_1^* \widetilde{X}_1+q_3^* \widetilde{X}_3} (x) \notag\\ &=& \frac{1}{q_3^*} (f_{\langle q^*,X \rangle+ q_1^* \widetilde{X}_1}(x) - f_{\langle q^*,X \rangle+ q_1^* \widetilde{X}_1+q_3^* \widetilde{X}_3} (x))\notag \\
 &=& \frac{1}{q_3^*} (f_{\langle q^*,X \rangle+ q_1^* \widetilde{X}_1+ q_2^* \widetilde{X}_2}(x) - f_{\langle q^*,X \rangle+ q_1^* \widetilde{X}_1+q_3^* \widetilde{X}_3} (x))\notag \\
 &=&\frac{q_3^*-q_2^*}{q_3^*} f^{'}_{\langle q^*,X \rangle+ q_1^* \widetilde{X}_1+q_2^* \widetilde{X}_2+q_3^* \widetilde{X}_3} (x) \label{contra1}
\end{eqnarray}
But from lemma \ref{unimodal} and \eqref{mod}, $f^{'}_{\langle q^*,X \rangle+ q_1^* \widetilde{X}_1+q_2^* \widetilde{X}_2}$ is strictly positive on $(0,x)$, thus
\begin{eqnarray}
 f^{'}_{\langle q^*,X \rangle+ q_1^* \widetilde{X}_1+q_2^* \widetilde{X}_2+q_3^* \widetilde{X}_3} (x) &=&f^{'}_{\langle q^*,X \rangle+ q_1^* \widetilde{X}_1+q_2^* \widetilde{X}_2}\star f_{q_3^* \widetilde{X}_3} (x) >0 \label{contra2}
\end{eqnarray}
Therefore, if $q_3^*$ is not equal to $q_1^*$, $q_2^*$ or 0, we must have $ f^{'}_{\langle q^*,X \rangle+ q_1^* \widetilde{X}_1+q_3^* \widetilde{X}_3} (x) =0$, in order to satisfy the KT conditions, but this contradicts \eqref{contra1} and \eqref{contra2}.
\\

\noindent
We have just shown that the KT conditions for minima can be satisfied only with points in $\Theta(n)$ that contains at most two different non-zero values, i.e. a minimizer has the following form (up to permutations) $$q^*=(\underbrace{p_1,\ldots,p_1}_{k},\underbrace{p_2,\ldots,p_2}_{l},0,\ldots,0),$$
with $k,l,k+l \in \{0,\ldots,t\}$, $p_1,p_2\in[0,1]$, such that $kp_1+lp_2=1$.
%We denote $f_{n_1,n_2} := f_{\langle(\underbrace{p_1,\ldots,p_1}_{n_1},\underbrace{p_2,\ldots,p_2}_{n_2}),X_{n_1+n_2} \rangle}$ with $n_1,n_2 \in \mN$, $X_{n_1+n_2}$ of size $n_1+n_2$ filled in with i.i.d copies of $X_1$ and $p_1$, $p_2$ being values of the point $q_c$ we are dealing with. \\
%As $q_c$ satisfies the KT conditions, we  have that
%$$\nabla_q \pp \{ \langle q,X \rangle \leq x\} \big|_{q=q_c} = cst,$$
%which we know from previous expansions to be equivalent to
%$$f_{k+1,l+1}^{'}(x)=0.$$
Let us assume that $k \geq 2$. We define $q^*_{\delta}:=q^* +\delta e_1 - \delta e_{2}$, with $0<\delta < p_1$ and $e_i \in \mR^t$ s.t. $(e_i)_j=\delta_{ij}$.
Since $q^*$ is a minimizer, we have
\begin{eqnarray}
 f^{'}_{\langle q^*,X \rangle + p_1\widetilde{X}_1+p_2 \widetilde{X}_2 }(x)=0
\end{eqnarray}
and using lemmas \ref{lemma_trick} and \ref{lemma_trick2}, we get
\begin{eqnarray}
\frac{\partial^2}{\partial \delta^2} \Big|_{\delta=0} \pp \{ \langle q^*_{\delta},X \rangle \leq x\} = 2 f^{'}_{\langle q^*,X \rangle + p_1\widetilde{X}_1+p_1 \widetilde{X}_2 } (x).
\end{eqnarray}
From the expansions in \eqref{contra1} and \eqref{contra2}, if $p_1<p_2$, we get $\frac{\partial^2}{\partial \delta^2} \Big|_{\delta=0} \pp \{ \langle q^*_{\delta},X \rangle \leq x\} < 0$, and $q^*$ cannot be a minimizer.
Thus the minimal component in $q^*$ has to appear only once. Say $p_1$ appears only once and $p_2$ appears $l$ times ($1\leq l \leq t-1$) and is greater than $p_1$, i.e. $p_2=\frac{1-p_1}{l}>p_1$, which implies $p_1<\frac{1}{l+1}$.
%Thus a minima having two non-zero different component can take place only for $p_1\in(0,\frac{1}{l+1})$.
At $p_1=\frac{1}{l+1}$, all components of $p^*$ are equal, and the function $p_1 \mapsto \pp \{ \langle q^*,X \rangle \leq x\}$ has an extremum at that point. Let us assume that there is at most one extremum within $(0,\frac{1}{l+1})$. A simple computation shows that $\frac{\partial}{\partial p_1} \Big|_{p_1=0} \pp \{ \langle q^*,X \rangle \leq x\} = -\frac{1}{l} f^{'}_{\frac{1}{l} \sum_{i=1}^{l+1} X_i} (x)$, which is strictly negative if and only if $x <1$, and
$\frac{\partial^2}{\partial p_1^2} \Big|_{p_1=\frac{1}{l+1}} \pp \{ \langle q^*,X \rangle \leq x\} = c(l) e^{-(l+1)x} (l+2-(l+1)x) x^{l+1}$, with $c(l)>0$, leading to a strictly positive second derivative if and only if $x < \frac{l+2}{l+1}$. Since we assumed that there is a most one extremum in $(0,\frac{1}{l+1})$, previous conditions on the sign of the first and second derivatives of $p_1 \mapsto \pp \{ \langle q^*,X \rangle \leq x\}$ at the boundaries of $(0,\frac{1}{l+1})$ imply that there is no minima that can occur in that interval. 

%We are thus left with showing that the densities $f_p$ of $p (X_1+\tilde{X}_1)+ \frac{1-p}{l} (\sum_{i=2}^{l+1} X_i+\tilde{X}_2)$ cannot
%have their maxima at the same location for $p\neq \tilde{p}$ and $p,\tilde{p} \in (0,\frac{1}{l+1})$.

So we need to show that $p_1 \mapsto \pp \{ \langle q^*,X \rangle \leq x\}$ has at most one extremum within $(0,\frac{1}{l+1})$, where $q^*=(p_1,\underbrace{\frac{1-p_1}{l},\ldots,\frac{1-p_1}{l}}_{l \text{ times}},0,\ldots,0)$. 
%We know that
%$$\frac{\partial^2}{\partial p_1^2} \pp \{ \langle q^*,X \rangle \leq x\}  = ( p_1 - \frac{1-p_1}{l}) f^{'}_{\langle q^*,X \rangle + p_1\widetilde{X}_1+  \frac{1-p_1}{l} \widetilde{X}_2 } (x).$$
We now use $p$ instead of $p_1$ and $k$ instead of $l$.
Let us define
$$f_{p,I,J}(x) = f_{ p \sum_{i=1}^I X_i + \frac{1-p}{k} \sum_{j=1}^J Y_i} (x), \quad x \in \mR, \, p \in (0,1), \,k,I,J \in \mZ_+, $$
where ${\{X_i\}}_{1\leq i \leq I}, {\{Y_j\}}_{1\leq k \leq J}\stackrel{iid}{\sim} \mathcal{E}(1)$.
We want to show a contradiction between the following assumptions:
\begin{eqnarray}
&& k \geq 1 \label{contrad1}\\
&& p,q \in (0,\frac{1}{k+1}), \, p \neq q \label{contrad2}\\
&& f_{p,2,k+1}^{'}(x)=f_{q,2,k+1}^{'}(x)=0, \, x\in \mR_+ \label{contrad3}
\end{eqnarray}
Since we are now working with simpler combination of our exponential random variables, we compute directly the convolution without using the Fourier transform. We have 
\begin{eqnarray}
 f_{p,2,k+1}^{'}(x) &=&(k+1)\left(\frac{1}{p}\right)^2 \left(\frac{1}{\bar{p}}\right)^{k+1} e^{-x/\bar{p}} \left(\frac{-1}{\Delta}\right)^{k+1} \\&& \left[ \left( \frac{\Delta p-\Delta x}{k+1} -1 \right) \left( e^{-\Delta x}- \sum_{l=0}^k \frac{(-\Delta x)^l}{l!} \right) + \frac{(-\Delta x)^{k+1}}{(k+1)!} \right] \notag
\end{eqnarray}
where
$$\Delta =\Delta (p)= \frac{1}{p}-\frac{1}{\bar{p}}, \quad \bar{p} = \frac{1-p}{k}.$$
%Hence, $f_{p,2,k+1}^{'}(x)=0$ is equivalent to
%\begin{eqnarray}
%\left( \frac{\Delta p-\Delta x}{k+1}  -1 \right) \left( e^{-\Delta x}- \sum_{l=0}^k \frac{(-\Delta x)^l}{l!} %\right) + \frac{(-\Delta x)^{k+1}}{(k+1)!} =0 .\label{cond0}
%\end{eqnarray}
Define
\begin{align}
\notag
g_k(\Delta,x) & \triangleq f'_{p,2,k+1}(x),\\ \notag
c_k(\Delta, x) & \triangleq (k+1) \left( \frac{1}{p} \right)^2 \left( \frac{1}{\bar{p}} \right)^{k+1} e^{-x/\bar{p}} \left( \frac{-1}{\Delta} \right)^{k+1},\\ \notag
t_k(\Delta, x) & \triangleq \left( e^{-\Delta x} - \sum_{l = 0}^k \frac{(-\Delta x)^l}{l!} \right),
%\gamma (\Delta) & \triangleq {\operatorname{d} \Delta p \over \operatorname{d} \Delta} = \frac{p^2k}{(1-p)^2+p^2k},
\end{align}
and let $\x(\Delta)$ be the solution of $g_k(\Delta,\x(\Delta)) = 0$. Thus,
\begin{equation}\label{xproperty}
-\left( \frac{\Delta p-\Delta \x}{k+1} -1 \right) \left( e^{-\Delta \x} - \sum_{l = 0}^k \frac{(-\Delta \x)^l}{l!} \right) = \frac{(-\Delta \x)^{k+1}}{(k+1)!}.
\end{equation}
Then
\begin{equation}\label{dg/dD}
0 = {\operatorname{d} g_k(\Delta,\x(\Delta)) \over \operatorname{d} \Delta} = {\partial g_k(\Delta,x) \over \partial \Delta} \Bigg|_{x = \x(\Delta)} +  {\operatorname{d} \x(\Delta) \over \operatorname{d} \Delta} \cdot {\partial g_k(\Delta,x) \over \partial x} \Bigg|_{x = \x(\Delta)}
\end{equation}
%From~\eqref{xproperty}, 
%We can calculate ${\partial g_k(\Delta,x) \over \partial \Delta} \Bigg|_{x = \x(\Delta)}$ as
and
\begin{align}
\notag
&{\partial g_k(\Delta,x) \over \partial \Delta} \Bigg|_{x = \x(\Delta)}\\ \label{eq7}
& = c(\Delta,x^*) \Bigg{[} \bigg( \frac{\gamma (\Delta)- x^*}{k+1} \bigg) \Bigg( e^{-\Delta x^*} - \sum_{l = 0}^k \frac{(-\Delta x^*)^l}{l!} \Bigg) \\ \notag
& + (-x^*) \bigg( \frac{\Delta p-\Delta x^*}{k+1} -1 \bigg) \Bigg( e^{-\Delta x^*} - \sum_{l = 0}^{k-1} \frac{(-\Delta x^*)^l}{l!} \Bigg) + (-x^*) \frac{(-\Delta x^*)^k}{k!} \Bigg{]} \\% \label{eq8}
%& = c(\Delta,x^*) \Bigg{[} \bigg( \frac{\gamma (\Delta)- x^*}{k+1} \bigg) \Bigg( e^{-\Delta x^*} - \sum_{l = 0}^k \frac{(-\Delta x^*)^l}{l!} \Bigg) \\ \notag
%& + (-x^*) \bigg( \frac{\Delta p-\Delta x^*}{k+1} -1 \bigg) \Bigg( e^{-\Delta x^*} - \sum_{l = 0}^{k} \frac{(-\Delta x^*)^l}{l!} + \frac{(-\Delta x^*)^k}{k!} \Bigg) + (-x^*) \frac{(-\Delta x^*)^k}{k!} \Bigg{]} \\ \label{eq9}
%& = c(\Delta,x^*) \Bigg{[} \bigg( \frac{\gamma (\Delta)- x^*}{k+1} \bigg) t(\Delta,x^*) + (-x^*) \bigg( \frac{\Delta p-\Delta x^*}{k+1} -1 \bigg) t(\Delta,x^*) + (p-x^*) \frac{(-\Delta x^*)^{k+1}}{{k+1}!} \Bigg{]} \\ 
\label{eq10}
& = c(\Delta,x^*) \Bigg{[} \bigg( \frac{\gamma (\Delta)- x^*}{k+1} \bigg) t(\Delta,x^*) + (-x^*) \bigg( \frac{\Delta p-\Delta x^*}{k+1} -1 \bigg) t(\Delta,x^*) \\ \notag
& - (p-x^*) \bigg( \frac{\Delta p-\Delta x^*}{k+1} -1 \bigg) t(\Delta,x^*) \Bigg{]} \\ \label{eq11}
%& = c(\Delta,x^*) t(\Delta,x^*) \Bigg{[} \bigg( \frac{\gamma (\Delta)- x^*}{k+1} \bigg) - p \bigg( \frac{\Delta p-\Delta x^*}{k+1} -1 \bigg) \Bigg{]} \\ \label{eq12}
& = c(\Delta,x^*) t(\Delta,x^*) \bigg( -\frac{pk}{(1-p)(k+1)} \bigg) \Bigg( x^* - \bigg( 1 + \frac{p(1-p)}{(1-p)^2+p^2k} \bigg) \Bigg)
\end{align}
where~(\ref{eq7}) is from the definition of $x^*$, and~(\ref{eq10}) is from~(\ref{xproperty}).

%\begin{equation} \label{pg/pD}
%{\partial g_k(\Delta,x) \over \partial \Delta} \Bigg|_{x = \x(\Delta)} = c(\Delta,\x) t(\Delta,\x) \left( -\frac{pk}{(1-p)(k+1)} \right) \left( \x - \left( 1 + \frac{p(1-p)}{(1-p)^2+p^2k} \right) \right).
%\end{equation}
Similarly,
\begin{align}
& {\partial g_k(\Delta,x^*) \over \partial x} %\label{eq3-1} 
%& = c(\Delta,x^*) \Bigg{[} \bigg( \frac{-\Delta}{k+1} \bigg) \Bigg( e^{-\Delta x^*} - \sum_{l = 0}^k \frac{(-\Delta x^*)^l}{l!} \Bigg) \\ \notag
%& + (-\Delta) \bigg( \frac{\Delta p-\Delta x^*}{k+1} -1 \bigg) \Bigg( e^{-\Delta x^*} - \sum_{l = 0}^{k-1} \frac{(-\Delta x^*)^l}{l!} \Bigg) + (-\Delta) \frac{(-\Delta x^*)^k}{k!} \Bigg{]} \\ \label{eq3-2}
%& = c(\Delta,x^*) \Bigg{[} \bigg( \frac{-\Delta}{k+1} \bigg) t(\Delta,x^*) + (-\Delta) \bigg( \frac{\Delta p-\Delta x^*}{k+1} -1 \bigg) t(\Delta,x^*) + \bigg( \frac{\Delta p- \Delta x^*}{x^*} \bigg) \frac{(-\Delta x^*)^{k+1}}{{k+1}!} \Bigg{]} \\ \label{eq3-3}
%& = c(\Delta,x^*) \Bigg{[} \bigg( \frac{-\Delta}{k+1} \bigg) t(\Delta,x^*) + (-\Delta) \bigg( \frac{\Delta p-\Delta x^*}{k+1} -1 \bigg) t(\Delta,x^*) \\ \notag
%& - \bigg( \frac{\Delta p-\Delta x^*}{k+1} -1 \bigg) \bigg( \frac{\Delta p- \Delta x^*}{x^*} \bigg) t(\Delta,x^*) \Bigg{]} \\ \label{eq3-4}
%& = c(\Delta,x^*) t(\Delta,x^*) \Bigg{[} \bigg( \frac{-\Delta}{k+1} \bigg) - \bigg( \frac{\Delta p-\Delta x^*}{k+1} -1 \bigg) \frac{\Delta p}{x^*}  \Bigg{]} \\ 
 = c(\Delta,x^*) t(\Delta,x^*) \Bigg( -\frac{\Delta pk}{(1-p)(k+1)} \Bigg) \Bigg (\frac{x^*-1}{x^*} \Bigg) \label{pg/px}
\end{align}
%\begin{equation} \label{pg/px}
%{\partial g_k(\Delta,\x) \over \partial x} = c(\Delta,\x) t(\Delta,\x) \left( -\frac{\Delta pk}{(1-p)(k+1)} \right) \left( \frac{\x-1}{\x} \right).
%\end{equation}
Note that $c(\Delta,\x)\cdot t(\Delta,\x) > 0$ for all $p \in (0,\frac{1}{k+1})$, and~\eqref{dg/dD} can be written as
\begin{equation}\label{dg/dD_2}
0 = \left( \x - \left( 1 + \frac{p(1-p)}{(1-p)^2+p^2k} \right) \right) + \Delta \left( \frac{\x-1}{\x} \right) \cdot {\operatorname{d} \x(\Delta) \over \operatorname{d} \Delta}.
\end{equation}
For each $p$ and from Lemma \ref{unimodal}, there exists a single $x^*$ satisfying~\eqref{xproperty}, and from~\eqref{pg/px}, $\x \geq 1$.

%For each $p$, if there exists more than one $\x$ satisfying~\eqref{xproperty}, since $f_{p,2,k+1}^{'}(0) > 0$ and $f_{p,2,k+1}^{'}(x)<0$ for all large enough $x$, there must exist at least three zeros, $\x_1<\x_2<\x_3$, of $f_{p,2,k+1}^{'}(x)$, such that $f_{p,2,k+1}^{''}(\x_1) \leq 0, f_{p,2,k+1}^{''}(\x_2) \geq 0, f_{p,2,k+1}^{''}(\x_3) \leq 0$. However, from~\eqref{pg/px} we know that $\x_1 \geq 1$, which implies $\x_2 > 1$ and $f_{p,2,k+1}^{''}(\x_2) < 0$, hence a contradiction. Thus, there exists only one $\x$ satisfying~\eqref{xproperty}.\\
%\\
Suppose that there exists $p \in (0,\frac{1}{k+1})$ such that $\x(\Delta(p)) \leq 1 + \frac{p(1-p)}{(1-p)^2+p^2k}$ and let 
$$\p = \sup_{p \in (0,\frac{1}{k+1}), \ \x(\Delta(p)) \leq ( 1 + \frac{p(1-p)}{(1-p)^2+p^2k} )} p,$$
such that $\x(\Delta(p)) > ( 1 + \frac{p(1-p)}{(1-p)^2+p^2k} )$ for all $p \in (\p, \frac{1}{k+1})$ and $\x(\Delta(\p)) = ( 1 + \frac{\p(1-\p)}{(1-\p)^2+{\p}^2k} )$. 
Observe that 
%$\lim_{p \rightarrow \frac{1}{k+1}} \x = 1+ \frac{1}{k+1}, \ \lim_{p \rightarrow \frac{1}{k+1}} f_{p,2,k+1}^{'}(\x) = 0$, and 
${\operatorname{d} ( 1 + \frac{p(1-p)}{(1-p)^2+p^2k} ) \over \operatorname{d} p } > 0$ for all $p \in (0,\frac{1}{k+1})$, hence ${\operatorname{d} \x(\Delta(\p)) \over \operatorname{d} p } \geq {\operatorname{d} ( 1 + \frac{p(1-p)}{(1-p)^2+p^2k} ) \over \operatorname{d} p } |_{p=p^*} > 0$. However, from~\eqref{dg/dD_2}, we must have ${\operatorname{d} \x(\Delta(\p)) \over \operatorname{d} \Delta}={\operatorname{d} \x(\Delta(\p)) \over \operatorname{d} p } = 0$. This contradiction shows that $\x(\Delta(p)) > 1 + \frac{p(1-p)}{(1-p)^2+p^2k}$ for all $p \in (0,\frac{1}{k+1})$. Therefore, ${\operatorname{d} \x(\Delta) \over \operatorname{d} \Delta} < 0$ and ${\operatorname{d} \x(p) \over \operatorname{d} p} > 0$ for all $p \in (0,\frac{1}{k+1})$, and~\eqref{contrad3} can not hold given~\eqref{contrad1} and~\eqref{contrad2}.
\end{proof}

\section{Discussion}
There are at least two possible directions to generalize this conjecture. One is to consider the matrix version as stated in \cite{telatar} (partial results have been achieved for small rates but the general conjecture is still open).  
 Another direction is to investigate for what kind of distributions of the $X_i$'s Conjecture \ref{miso} may still hold. Given a sequence of i.i.d.\ random variables, how do we construct a weighted sum of them in order to minimize the probability of exceeding a given threshold? For exponential distributions, which may arise naturally in different applications (\emph{e.g.}, capacities, waiting times), the symmetry in the problem is not sufficient to lead to a fully symmetric solution (full diversity). But the minimizer still has a symmetric structure and a more general result leading to such minimizer would be interesting. Yet, our proof is only weakly relying on the symmetry of the problem. 
One can look at other examples. In the case of the Cauchy distribution, the function $q \mapsto \pp \{ \langle q,X \rangle \leq x\}$ is constant. In the case of the Gaussian distribution, the conjecture holds and $k$ can easily be determined.
\begin{lemma}
Let $n \in \mN^*$, $q \in \Theta(t):=\{ q \in \mR_+^t | \sum_{i=1}^t q_i =1\}$, $X=(X_1,\ldots,X_n)$ with ${(X_i)}_{1\leq i \leq n} \stackrel{iid}{\sim} \mathcal{N}(0,1)$. For $x >0$, $$\arg\min_{q \in \theta(n)} \pp \{\langle q,X \rangle \leq x\}\ni (1,0,\ldots,0),$$
for $x<0$, $$\arg\min_{q \in \theta(n)} \pp \{\langle q,X \rangle \leq x\}=(1/n,\ldots,1/n)$$
and for $x=0$, $\pp \{\langle q,X \rangle \leq x\}=1/2$, $\forall q \in \Theta(n)$. 
\end{lemma}
The first part of the proof, covered by Lemmas \ref{lemma_trick}, \ref{lemma_trick2} and \ref{unimodal}, could possibly be generalized to other distributions, by imposing conditions on the derivative of the Fourier transform of the sum (or of $X_1$), such that the KT conditions would only be satisfied under some symmetry of the $q_i$'s.
On the other hand, without any conditions on the random variables $\{X_i\}$, except to be i.i.d., the conjecture does not hold, i.e., a statement such as in Conjecture \ref{miso} is not universal. In fact, for $n=2$, $X_1,X_2 \stackrel{iid}{\sim} \mathcal{E}(1)$, $x=1.1$, the input that {\it maximizes} the outage probability is not of the form $(1,0)$, $(0,1)$ or $(1/2,1/2)$ (it is around (0.2,0.8)). Thus by choosing $Y_1,Y_2\stackrel{iid}{\sim}-X_1$ and $y=-1.1$ we get a counter-example to the conjecture if stated for any independent real random variables. Even if stated for positive random variables, one can consider $Z_1,Z_2\stackrel{iid}{\sim}L-X_1 \mathds{1}_{[0,L]}$ for a large enough $L \in \mR_+$, $z=2L-1.1$ and get a counter-example for positive random variables. Hence, one would have to find a more specific condition (than i.i.d.) under which Conjecture \ref{miso} may hold. 
Finally, note that Conjecture \ref{angle} reduces to a conjecture stated in terms of a weighted sum of random variables (as in Conjecture \ref{miso}) only when $H$ is unitary invariant. And the only way to have unitary invariance and independence is to assume a Gaussian distribution for $H$, which means that, the previous counter-examples to Conjecture \ref{miso} for arbitrary i.i.d.\ random variables do not lead to counter-examples to Conjecture \ref{angle} when $H$ has an arbitrary unitary invariant distribution.

\end{document}